\documentstyle[twoside,fleqn,espcrc2]{article}
\newcommand\ra{\Rightarrow}

\newcommand\bra{\langle}
\newcommand\ket{\rangle}

\newcommand\um{\frac12}
\newcommand\la{\lambda}
\newcommand\C{{\cal M}}
\newcommand{\tp}{\tilde{\varphi}}
\newcommand{\ba}{\begin{array}}
\newcommand{\bb}{\Box}

\newcommand{\ea}{\end{array}}
\newcommand{\eq}{\begin{equation}}
\newcommand{\en}{\end{equation}}

\newcommand{\ZZ}{\hbox{{\rm Z{\hbox to 3pt{\hss\rm Z}}}}}

\newcommand{\AmS}{{\protect\the\textfont2
  A\kern-.1667em\lower.5ex\hbox{M}\kern-.125emS}}
\title{The dual sector of
 the $\lambda\phi^4$ Theory in 4D
\thanks{presented by F. Gliozzi. This work is supported in part by
M.U.R.S.T.}}
\author{F. Gliozzi and  M. Leone \address{Dipartimento di Fisica
Teorica, Universit\`a
di Torino, \\
  via P. Giuria 1, 10125 Torino, Italy} }
\begin{document}
\begin{abstract}
The one-component $\lambda\phi^4$ theory in four dimensions
 in the spontaneously broken symmetry phase
 has a non-trivial non-perturbative sector which can be studied by means of
 a duality transformation of its Ising limit.
 Duality maps this theory to a model of interacting membranes.
  Physical states correspond to membrane excitations.
 The way this non-perturbative behaviour can be reconciled with the
 triviality of the theory in its continuum limit is discussed.
\end{abstract}
\maketitle
\section{INTRODUCTION}
One of the most powerful sources of non-perturbative information on
a quantum field theory is duality.  A typical duality transformation is
an exact one-to-one map between two different descriptions of the same
physical system, such that a strong coupling region of a description
corresponds to a weak coupling region of the other. One of the simplest
examples is the Kramers-Wannier duality  of the Ising model \cite{kw}.
The two-dimensional Ising model in a square lattice is self-dual, $i.e.$
the above-mentioned descriptions differ simply by the value of the coupling
constant, while the dual description of the three-dimensional Ising model
is  a $\ZZ_2$  gauge theory.
 Although it is  known that the Kramers-Wannier transformation can be
extended to the Ising model defined on a  lattice of arbitrary dimension $D$, there has been apparently no systematic study for $D\ge4$. For $D=4$ preliminary
results have been reported in Lattice'97 conference\cite{fg}. Here we report a
more complete analysis.
The main consequence of this duality transformation is that the physical
spectrum of this theory necessarily contains, in the
broken symmetry phase, besides the fundamental scalar particle, a
(possibly) infinite tower of massive states of any spin  and parity.
Since this Ising model belongs to
the universality class of the one-component $\phi^4$-theory,
it is most likely trivial, $i.e.$ its continuum limit
should flow to an infrared Gaussian fixed point, hence the excited states
should disappear in the free field limit. This can be obtained assuming
that the ratios between the masses
of the excitations and of the fundamental particle go to infinity, as our
numerical data seem to indicate. As a consequence, the power law obeyed by
these ratios defines a new non-trivial exponent of $\phi^4_4$ universality
class.
It is curious that the request of triviality,
when combined with duality, implies the existence of non-trivial directions
in the RG flow.
This could be related to the observed non trivial directions in scalar
theories with non-polynomial potentials \cite{kh}, indeed the correlators
we use to extract the excited spectrum are non-local and non-polynomial
functions of $\phi$.
\section{DUAL DESCRIPTION}
The $4D$ Ising model on the hypercubic lattice  ${\cal L}= \ZZ^4$ defined
by the action
\eq
S_{links}=-\sum_{x\in{\cal L}}\sum_{\mu=1}^4\phi_x\phi_{x+\hat{\mu}} ~~,
\label{links}
\en
where $\phi_x=\pm1$ is the {\sl order} operator associated to the sites
$x\in{\cal L}$,
admits a dual description in terms of a dual field $\tp_\bb=\pm1$
associated to the plaquettes of the dual lattice
$\tilde{\cal L}= (\ZZ+\um)^4$.
The dual action is given by the sum of the contributions of the elementary
cubes of $\tilde{\cal L}$:
\eq
S_{cubes}=\sum_{cubes\in\tilde{\cal L}} \tp_{cube}~,~~\tp_{cube}=
\prod_{\bb\in cube} \tp_\bb~~.
\en
Kramers-Wannier duality states that the partition function
$Z_{Ising}(\beta)=\sum_{\{\phi_x\}}\exp(-\beta S_{links}) $ is proportional to
the partition function of the dual
description $Z_{dual}(\tilde{\beta}) =\sum_{\{\tp_\bb\}}
\exp(-\tilde\beta S_{cubes})$~, provided that their couplings are related by
$\sinh(2\beta)\sinh(2\tilde{\beta})=1$.
This shows in particular that the low temperature region of the Ising model
is mapped into the strong coupling region of the dual model.
 Such a dual description has a local $\ZZ_2$  symmetry generated by
 any arbitrary function $\eta_{link}=\pm1$ of the links of
 $\tilde{\cal L}$, through the transformation \cite{fg}
 \eq
 \tp_\bb\to\tp_\bb'=\eta_\bb\,\tp_\bb ~~,~~
 \eta_\bb=\prod_{links\in\bb}\eta_{link} ~~.
 \label{sym}
 \en
 The continuum version of this theory is known as Kalb-Ramond model
 \cite{kr}
 %This is the lattice version of the generalized gauge transform of
 %an antisymmetric two-index potential   $A_{\mu\nu}\sim\tp_\bb$, i.e.
 %A_{\mu\nu}\to A_{\mu\nu}+\pa_\mu\eta_\nu-\pa_\nu\eta_\mu$
 %which generates the gauge invariant three-index field strength
 %$F_{\mu\nu\rho}\sim\tp_{cube}$.
 The local symmetry (\ref{sym}) implies that the
 {\sl disorder} observables \cite{kc}, {\sl i.e.} the  invariant
 quantities of the dual description, are the vacuum expectation values of
 (products of) {\sl surface operators}  $\tp_\Sigma$  associated to any
 arbitrary, closed surface $\Sigma$ of $\tilde{\cal L}$:
 \eq
 \tp_\Sigma=\prod_{\bb\in \Sigma}\tp_\bb~~.
 \en
 In the phase corresponding to the broken $\ZZ_2$ symmetry of the
 Ising model, the connected correlator between surface operators
 $\bra \tp_\Sigma\,\tp_{\Sigma'}\ket_{c}$ has a
 strong coupling expansion which can be expressed
  as a sum of weighted 3D  manifolds bounded by
$\Sigma$ and $\Sigma'$ . These $3D$ manifolds can be viewed as
the world volumes swept out by a close membrane in its 'time' evolution
from $\Sigma$ to $\Sigma'$. Then,
the dual theory may be reformulated as a model of interacting
2-branes, in the same sense as the strong
 coupling expansion of any lattice gauge theory can be seen as a string
 theory. As a consequence, the physical spectrum of this theory may
 be associated to membrane excitations, in the same way as the glue-ball
 spectrum of the gauge theories can be associated to closed string
 states.
 \subsection{ The membrane spectrum }
 In order to study the physical spectrum of the membrane one has to
 define suitable correlators between surface operators with a very
 simple recipe:
$i)$ choose a (possibly) large set of closed surfaces $\{\Sigma_1,\Sigma_2,
\cdots,\Sigma_n\}$ belonging to a given $3D$ slice ${\cal S}(x_4)$ of
the dual lattice, where $x_4$ plays  the role of  ``time''
coordinate;
$ii)$ evaluate the {\sl connected} correlation matrix
\eq
C_{ij}(t)=\bra \sum_{x\in{\cal S}(x_4)}\tp_{\Sigma_i}
\sum_{x\in{\cal S}(x_4+t)}\tp_{\Sigma_j}\ket_{c} ~~;
\label{cor}
\en
$iii)$ project on states of definite spin and parity.
The sum over $x$ in Eq.(\ref{cor}) is a shorthand notation for
the projection on zero momentum states.
For sufficiently large $t$ it is expected that the eigenvalues $\la_i$ of $C(t)$ have the
asymptotic form $\la_i \sim c_i e^{-m_it}$~,
where $m_i$ denotes the mass of the $i^{th}$ eigenstate.
This membrane model is an alternative formulation of the {\sl same} physical
system described by the Ising model, thus the spectrum of physical states must be the
same in both formulations. In particular, the mass of the ground membrane
state should coincide with that of the fundamental scalar particle of the
Ising model.
 In general, in a theory with order  and disorder operators, physical states
which couple to order operators are not necessarily coupled to the disorder
ones. This raises an obvious  question: is the lowest state coupled
to $\tp_\Sigma$ the same physical state  which couples to $\phi_x$?
for getting an answer one has to study also the mixed correlator
\eq
C_{io}(t)=\bra\sum_{\{\Sigma_i\in {\cal S}(x_4)\}}\tp_{\Sigma_i}
\sum_{x\in {\cal S}(x_4+t)\}}\phi_x\ket_{c}
\en
It turns out it is not vanishing (and { negative}),
then the ground particle couples both to order and disorder operators.
\begin{table*}[hbt]
% space before first and after last column: 1.5pc
% space between columns: 3.0pc (twice the above)
\setlength{\tabcolsep}{1.5pc}
% -----------------------------------------------------
% adapted from TeX book, p. 241
\newlength{\digitwidth} \settowidth{\digitwidth}{\rm 0}
\catcode`?=\active \def?{\kern\digitwidth}
% -----------------------------------------------------
\caption{The low lying mass spectrum }
\label{tab:masses}
\begin{tabular*}{\textwidth}{@{}l@{\extracolsep{\fill}}rrrr}
\hline
                 & \multicolumn{2}{l}{Lattice size $12^3\times16$}
               & \multicolumn{2}{l}{Lattice size  $16^4$} \\
\cline{2-3} \cline{4-5}
$J^P$                 & \multicolumn{1}{r}{mass at $\beta=0.154$}
                 & \multicolumn{1}{r}{$\#$ config.}
                 & \multicolumn{1}{r}{mass at $\beta=0.152$}
                 & \multicolumn{1}{r}{$\#$ config.}         \\
\hline
$0^+$   & $ 0.553(2)$ & $2.2~10^6$ & $ 0.383(3)$ & $ 2.0~10^6$ \\
$2^+$ & $ 2.06(4)$ & $2.1~10^6$ & $1.98(5)$        & $ 2.3~10^6$ \\
$(0^+)$'      & $1.42(6)$ & $1.0~10^6$ &  &  \\
$1^-$         & $ 1.8(2)$ & $2.1~10^6$ & $1.6(4)$        & $ 2.3~10^6$ \\
$3^-$           & $ 2.3(2)$ & $0.5~10^6$ &  &  \\
$0^-$         &  $2.6(6)$      & $0.5~10^6$    &    &  \\
\hline
\end{tabular*}
\end{table*}
The ground particle is associated  to the highest eigenvalue
$\lambda(t)$ of $C(t)$ and also to the highest eigenvalue
$\mu(t)$ of the mixed correlation matrix, denoted by ${\C}$.
We can write
\eq
\C(t)\vert\psi\ket=\mu(t)\sum_{i=1,n}
\psi_i\vert\tp_{\Sigma_i}\ket+\mu(t)\psi_o\vert\phi_x\ket
\en
The component $\psi_i$ of the highest eigenvector measures the
coupling of the ground particle to the source $\tp_{\Sigma_i}$
 The main features of the spectrum can be understood by assuming
that $\psi_i$ , apart an obvious normalization factor, does not depend
on the number $n$ of entries of the correlation matrices $C(t)$ and
$\C(t)$,but is only a function of the source $\Sigma_i$~:
 \eq
 \psi_i=\psi(\Sigma_i)~~.
\label{fach}
\en
This assumption  seems very well verified by the numerical data and has many
important consequences. In particular, one easily derives the following
equation, valid for any $t$
\eq
\mu^2-\mu(\lambda+C_{oo})+C_{oo}\lambda-\sum_j C_{jo}^2=0 ~,
\label{two}
\en
where $C_{oo}(t)$ is the order-order correlator.
It follows that $\C(t)$ has two eigenvalues $\mu(t)>0$  and $\bar\mu(t)<0$
(the highest and the lowest ones) with the same $t$ dependence,
 hence the ground particle is a degenerate doublet. Moreover, denoting with
$\vert\psi\ket$ and $\vert\bar\psi\ket$ the corresponding eigenvectors,
Eq. (\ref{fach}), when combined with the orthogonality
$\bra\psi\vert\bar\psi\ket=0$,
yields
$
\bar\psi_i=\psi_i~,~\bar\psi_o=-\psi_o/\sum_j\psi_j^2$~,
which shows that the two states of the doublet differ by the sign
of the coupling to the order parameter $\phi_x$.
Any other physical state $\vert\varphi\ket$ is decoupled from $\phi_x$ , because
\eq
\bra\psi\vert\varphi\ket=\bra\bar\psi\vert\varphi\ket=0~\ra~ \varphi_o=0~~.
\en
\section{NUMERICAL RESULTS}
In order to evaluate the connected correlations matrices and to extract the
low lying mass spectrum, we performed a series of numerical simulations
directly in the Ising model, using the method described in \cite{fg}.
The results are collected in Tab.1.
The values of $\beta$ and of lattice sizes are  those of a
large scale numerical simulation \cite{uw}, where  it was observed small
finite volume effects and a good agreement with 3-loop $\beta$-function.

The mass $m_{0^+}$ of the lowest membrane state coincides, within the
errors, with that of the fundamental particle contributing to the spin
correlation function \cite{uw}, as required by eq.(\ref{two}). The mass
spectrum has,
roughly, the pattern of the glue-ball spectrum of the $3D$ gauge
Ising model \cite{mi} (actually, when one of the $4D$ lattice
dimensions is squeezed, our dual model flows to a $3D~\ZZ_2$ gauge theory
and the membrane spectrum must flow to that of glue-balls). However
the ratio $m/m_{0^+}$ for the excited states, when available, is not a
constant, but  is rising when approaching the critical point at
$\beta_c=0.14967(3)$. This seems to agree with triviality
of $\phi^4_4$ theory, which suggests, when combined with duality,
that these ratios be universal functions of the renormalized coupling $g_r$
diverging at $g_r=0$.
The exponent of this divergence is presently unknown.

\end{document}